\documentclass[twocolumn,showpacs,preprintnumbers,amsmath,amssymb]{revtex4}

\usepackage{graphicx}
\usepackage{dcolumn}
\usepackage{bm}

\begin{document}

\preprint{Preprint submitted to JPSJ}

\title{Structures and Electromagnetic Properties of New Metal-Ordered Manganites; $R$BaMn$_{2}$O$_{6}$ ($R$ = Y and Rare Earth Elements)}

\author{Tomohiko Nakajima}%
\altaffiliation[e-mail: ]{t-nakaji@issp.u-tokyo.ac.jp}
\author{Hiroshi Kageyama}%

\author{Yutaka Ueda}%
\affiliation{%
Materials Design and Characterization Laboratory, Institute for Solid State Physics, University of Tokyo, 5-1-5 Kashiwanoha, Kashiwa, Chiba 277-8581, Japan}%

\author{Hideki Yoshizawa}
\affiliation{
Neutron Scattering Laboratory, Institute for Solid State Physics, The University of Tokyo, 106-1 Shirakata, Tokai, Ibaraki 319-1106, Japan}%

\date{\today}

\begin{abstract}
New metal-ordered manganites $R$BaMn$_{2}$O$_{6}$ have been synthesized and 
investigated in the structures and electromagnetic properties. 
$R$BaMn$_{2}$O$_{6}$ can be classified into three groups from the structural 
and electromagnetic properties. The first group ($R$ = La, Pr and Nd) has a 
metallic ferromagnetic transition, followed by an $A$-type antiferromagnetic 
transition in PrBaMn$_{2}$O$_{6}$. The second group ($R$ = Sm, Eu and Gd) 
exhibits a charge-order transition, followed by an antiferromagnetic long 
range ordering. The third group ($R$ = Tb, Dy and Ho) shows successive three 
phase transitions, the structural, charge/orbital-order and magnetic 
transitions, as observed in YBaMn$_{2}$O$_{6}$. Comparing to the 
metal-disordered manganites ($R^{3 + }_{0.5}A^{2 + }_{0.5})$MnO$_{3}$, 
two remarkable features can be recognized in $R$BaMn$_{2}$O$_{6}$; (1) 
relatively high charge-order transition temperature and (2) the presence of 
structural transition above the charge-order temperature in the third group. 
We propose a possible orbital ordering at the structural transition, that is 
a possible freezing of the orbital, charge and spin degrees of freedom at 
the independent temperatures in the third group. These features are closely 
related to the peculiar structure that the MnO$_{2}$ square-lattice is 
sandwiched by the rock-salt layers of two kinds, $R$O and BaO with extremely 
different lattice-sizes.
\end{abstract}

\maketitle

The magnetic and electrical properties of perovskite-type manganites with 
the general formula ($R^{3 + }_{1 - x}A^{2 + }_{x})$MnO$_{3}$ ($R$ = rare 
earth elements and $A$ = alkaline earth elements) have been extensively 
investigated for the last decade [1]. Among the interesting features are the 
so-called colossal magnetoresistance (CMR) and metal-insulator (M-I) 
transition accompanied by charge and orbital ordering. It is now widely 
accepted that these enchanting phenomena are caused by the strong 
correlation/competition of multi-degrees of freedom, that is, spin, charge, 
orbital and lattice.

The structure of perovskite $R$MnO$_{3}$ consists of MnO$_{2}$ 
square-sublattice and $R$O rock-salt-sublattice. The mismatch between the 
larger MnO$_{2}$ sublattice and the smaller $R$O sublattice is relaxed by 
tilting and rotating MnO$_{6}$ octahedra, leading to the lattice distortion 
from cubic to, mostly, orthorhombic GdFeO$_{3}$-type structure. At this 
lattice distortion, the bond angle $\angle $Mn-O-Mn deviates from 180\r{ }, 
resulting in a significant change of an effective one-electron bandwidth 
($W)$ or equivalently $e_{g}$-electron transfer interaction ($t)$. In the 
substitution system of ($R^{3 + }_{1 - x}A^{2 + }_{x})$MnO$_{3}$ with a 
fixed $x$ and a random distribution of $R^{3 + }$ and $A^{2 + }$, the structural 
and electromagnetic properties have been explained by the degree of 
mismatch, that is the tolerance factor $f$ = ($<$$r_{A}$$>$+$r_{\rm O})$/[$\sqrt{2}$($r_{\rm Mn}+r_{\rm O})$], where $<$$r_{A}$$>$, $r_{\rm Mn}$ and $r_{\rm O}$ are (averaged) ionic 
radii for the respective elements, because $W$ or $t$ is changed by varying $f$. 
\begin{figure}
\includegraphics{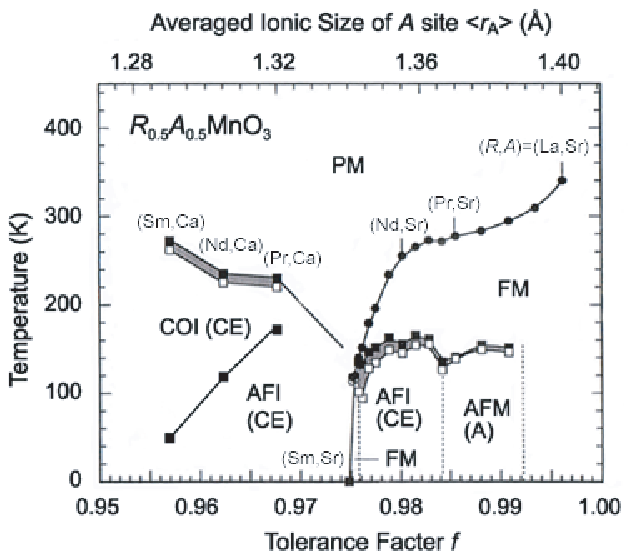}
\caption{\label{fig:fig1} Generalized phase diagram for ($R^{3 + }_{0.5}A^{2 + 
}_{0.5})$MnO$_{3}$ (Ref. 1). FM: ferromagnetic metal, AFM($A)$: $A$-type 
antiferromagnetic metal, AFI(CE): CE-type antiferromagnetic insulator, 
COI(CE): CE-type charger/orbital ordered insulator, PM: paramagnetic metal.}
\end{figure}

Figure ~\ref{fig:fig1} shows the generalized phase diagram for ($R^{3 + }_{0.5}A^{2 + 
}_{0.5})$MnO$_{3}$ expressed as a function of $f $[1], where the ferromagnetic 
metallic (FM) state due to the double-exchange (DE) interaction is dominant 
near $f \sim $ 1 (maximal $W$ or $t)$, while the CE-type charge/orbital-ordered 
(CO) state is most stabilized in the lower $f$ region ($f$ $<$ 0.975). In the middle 
region ($f \sim $ 0.975), the competition between the ferromagnetic DE and 
the antiferromagnetic CO interactions results in various phenomena including 
CMR.

Recently, it has been argued how the $A$-site randomness affect the physical 
properties of ($R^{3 + }_{1 - x}A^{2 + }_{x})$MnO$_{3}$. The phenomena 
such as the coexistence of FM phase with CO phase and the electronic phase 
separation [2] may come from the $A$-site randomness. Unfortunately, almost all 
the works devoted to a series of perovskite-type manganites so far are on 
the disordered perovskite-type manganites with $R^{3 + }$ and $A^{2 + }$ ions 
being randomly distributed. This means that, whenever $x$ is finite, there 
inevitably exists a disorder in the lattice. Since the physical properties 
of the manganite perovskite are quite sensitive to even a tiny change in 
lattice distortion, it is important to employ a compound without $A$-site 
disorder in order to make clear the effect of $A$-site randomness.

\begin{figure}
\includegraphics{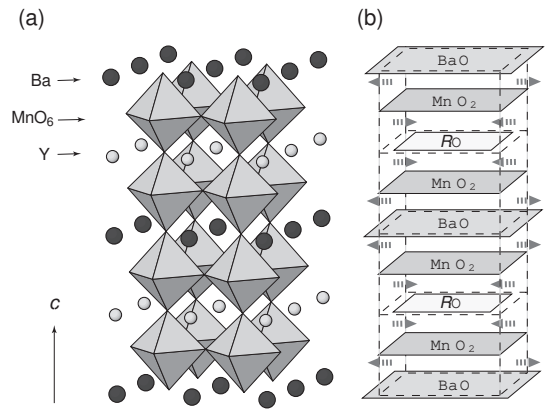}
\caption{\label{fig:fig2} Crystal structure of YBaMn$_{2}$O$_{6}$ (a) and a schematic illustration of structure for $R$BaMn$_{2}$O$_{6}$ (b). The MnO$_{2}$ 
square-lattice is sandwiched by the rock-salt layers of two kinds, $R$O and BaO with the different lattice-sizes.}
\end{figure}

Very recently, we successfully synthesized a metal-ordered perovskite-type 
manganite YBaMn$_{2}$O$_{6}$ with a successive stacking of 
YO-MnO$_{2}$-BaO-MnO$_{2}$-YO (see Fig.~\ref{fig:fig2}(a)) and observed successive three 
phase transitions; a structural transition without any charge and magnetic 
order at $T_{\rm S}$ = 520 K, a CO transition (M-I transition) at $T_{\rm CO}$ = 480 K 
and an antiferromagnetic transition at $T_{\rm N}$ = 195 K [3]. The observed 
$T_{\rm CO}$ = 480 K is the highest among the perovskite-type manganites. Across 
the phase transition at $T_{\rm S}$ = 520 K, the resistivity shows little change 
and the magnetic susceptibility exhibits a large reduction. Furthermore the 
magnetic interaction seems to be changed from ferromagnetic above $T_{\rm S}$ to 
antiferromagnetic below $T_{\rm S}$. Such transition was first observed in the 
perovskite manganites. The expectation that such novel transition could be 
closely related to the metal-ordered structure drove us to the study of 
metal-ordered perovskite-type manganites $R$BaMn$_{2}$O$_{6}$. In this paper we 
report the synthesis, structures and physical properties of new 
metal-ordered perovskite-type manganites $R$BaMn$_{2}$O$_{6}$ with a successive 
stacking of $R$O-MnO$_{2}$-BaO-MnO$_{2}$-$R$O. We summarize the results as a phase 
diagram and we compare the obtained phase diagram with that of ($R^{3 + 
}_{1 - x}A^{2 + }_{x})$MnO$_{3}$ shown in Fig. 1.

Powder samples were prepared by a similar solid-state reaction of 
$R_{2}$O$_{3}$, BaCO$_{3}$ and MnO$_{2}$ to that used for YBaMn$_{2}$O$_{6}$ 
[3]. The obtained products were checked to be single phases by X-ray 
diffraction. No perovskite-type compound was produced for Ce, Yb and Lu. The 
Er- and Tm-compounds included a significant amount of impurity phase.

The crystal structure was determined for 300-573 K by powder X-ray 
diffraction using CuK$\alpha $ radiation. The superlattice with a charge and 
orbital ordering was investigated by electron diffraction. The magnetic 
properties were studied using a SQUID magnetometer in a temperature range 
$T$ = 5-700 K under a magnetic field of 0.1 T. The electric resistivity of a 
sintered pellet was measured for $T$ = 100-620 K by a conventional four-probe 
technique.

The X-ray diffractions of all compounds clearly show the 
(0,0,1/2)-reflection indexed with the simple cubic perovskite structure, 
which is an evidence for the same metal-ordered structure as that of 
YBaMn$_{2}$O$_{6}$. The crystal structure at room temperature is tetragonal 
($a_{\rm p}$$\times$$a_{\rm p}$$\times$2$ c_{\rm p})$ in La- and Pr-compounds, while in the compounds 
with $R$ = Sm $\sim $ Ho it has a larger cell ($\sqrt{2}$$a_{\rm p}$$\times$$\sqrt{2}$$b_{\rm p}$$\times$2$ c_{\rm p})$ as observed in YBaMn$_{2}$O$_{6}$ [3], where $a_{\rm p}$, 
$b_{\rm p}$ and $c_{\rm p}$ denote the primitive cell for the simple cubic perovskite. 
The X-ray diffraction pattern of Nd-compound exhibits a mixture of 
($a_{\rm p}$$\times$$a_{\rm p}$$\times$2$ c_{\rm p})$- and ($\sqrt{2}$$a_{\rm p}$$\times$$\sqrt{2}$$b_{\rm p}$$\times$2$ c_{\rm p})$-phases 
at room temperature.

\begin{figure}
\includegraphics{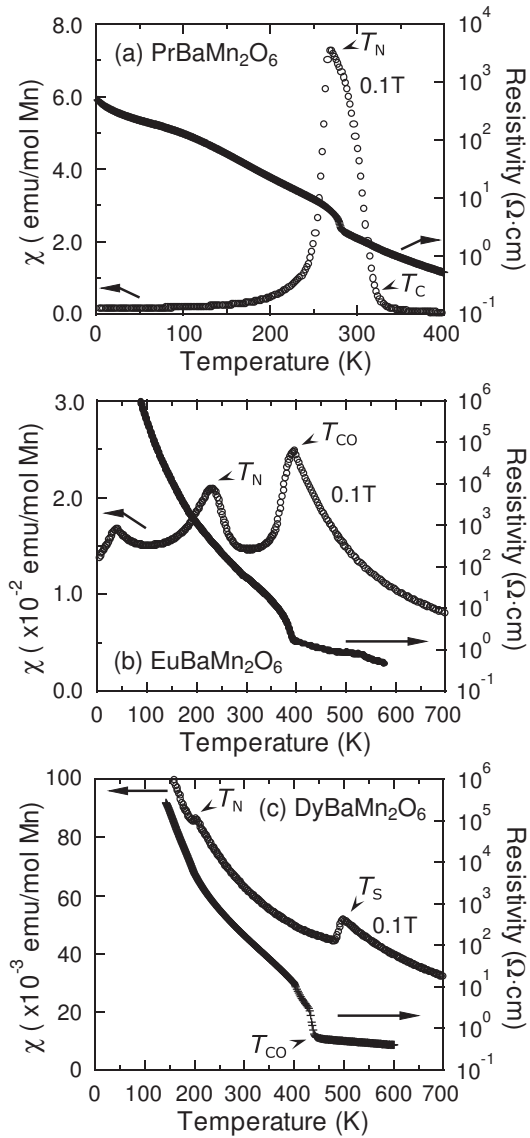}
\caption{\label{fig:fig3} Temperature dependence of resistivity and magnetic susceptibility 
(\textit{$\chi $}) for (a) PrBaMn$_{2}$O$_{6}$, (b) EuBaMn$_{2}$O$_{6}$ and (c) 
DyBaMn$_{2}$O$_{6}$. $T_{\rm C}$, $T_{\rm N}$, and $T_{\rm CO}$ represent the 
ferromagnetic, antiferromagnetic and charge/orbital-ordered transition 
temperatures, respectively. DyBaMn$_{2}$O$_{6}$ shows a structural 
transition without any charge and magnetic order at $T_{\rm S}$ (see the text).}
\end{figure}

$R$BaMn$_{2}$O$_{6}$ can be classified into three groups from the obtained 
structural and electromagnetic properties. The first group 
$R$BaMn$_{2}$O$_{6}$ with $R^{3 + }$ = La$^{3 + }$, Pr$^{3 + }$ and Nd$^{3 + }$ 
has a FM transition at $T_{\rm C}$, followed by antiferromagnetic transitions in 
PrBaMn$_{2}$O$_{6}$ (see Fig.~\ref{fig:fig3}(a)) and NdBaMn$_{2}$O$_{6}$. The obtained 
$T_{\rm C}$ for LaBaMn$_{2}$O$_{6}$ agrees well with the previous report [4]. The 
neutron magnetic diffraction study has revealed an $A$-type antiferromagnetic 
transition for PrBaMn$_{2}$O$_{6}$ [5]. This is consistent with the 
relatively low resistivity below $T_{N}$ compared with that in the CO state 
of EuBaMn$_{2}$O$_{6}$ or DyBaMn$_{2}$O$_{6}$, as shown in Fig. 3. A 
semiconductive behavior in the paramagnetic metallic (PM) state of 
PrBaMn$_{2}$O$_{6}$ is due to loosely sintered samples. The second group 
consists of $R$BaMn$_{2}$O$_{6}$ with $R^{3 + }$ = Sm$^{3 + }$, Eu$^{3 + }$ and 
Gd$^{3 + }$. The compounds exhibit CO transitions, followed by 
antiferromagnetic long range ordering. Figure 3(b) shows a typical example 
of magnetic susceptibility and resistivity for EuBaMn$_{2}$O$_{6}$. The 
third group includes the compounds with $R^{3 + }$ = Tb$^{3 + }$, Dy$^{3 + }$ 
and Ho$^{3 + }$ whose ionic radii are close to Y$^{3 + }$. These compounds 
show three phase transitions as observed in YBaMn$_{2}$O$_{6}$ [3]. The 
magnetic susceptibility and resistivity of DyBaMn$_{2}$O$_{6}$ are shown in 
Fig. 3(c) as an example. The distinct transitions at $T_{\rm S}$, $T_{\rm CO}$ and 
$T_{\rm N}$ are commonly observed in this series. The change of magnetic 
interaction from ferromagnetic above $T_{\rm S}$ to antiferromagnetic below 
$T_{\rm S}$ is not so clear in $R$BaMn$_{2}$O$_{6}$ ($R$ = Tb, Dy, Ho) as in 
YBaMn$_{2}$O$_{6}$ [3], because of the significant contribution of magnetic 
rare earth ions to the total magnetic susceptibility. However the reduction 
of magnetic susceptibility at $T_{\rm S}$ can be ascribed to a change of 
Weiss-temperature, namely a change of magnetic interaction, between the PM 
states above and below $T_{\rm S}$. 

The results are summarized in Fig. 4 as a phase diagram. Here we express the 
phase diagram as a function of the ratio of ionic radius, $r_{R^{3 + }}$/$r_{Ba^{2 + }}$ [6], instead of $f$. In $R$BaMn$_{2}$O$_{6}$, the MnO$_{2}$ 
sub-lattice is sandwiched by the rock-salt layers of two kinds, $R$O and BaO 
with the different lattice-sizes, as shown in Fig. 2(b) and therefore the 
tolerance factor cannot be defined. The ratio, $r_{R^{3 + }}$/$r_{Ba^{2 + }}$ is a 
measure of mismatch between $R$O- and BaO-lattices.

\begin{figure}
\includegraphics{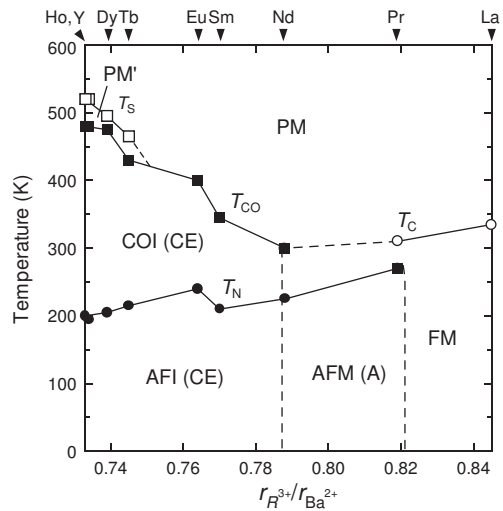}
\caption{\label{fig:fig4} Phase diagram for $R$BaMn$_{2}$O$_{6}$. The notation of each phase 
and transition temperature is the same as that defined in Fig.1 and Fig. 3 
(see the text). The PM' is a paramagnetic metal with a possible orbital ordering (see the text). Comparing to the ($R^{3 + }_{0.5}A^{2 + 
}_{0.5})$MnO$_{3}$, two remarkable features can be recognized in 
$R$BaMn$_{2}$O$_{6}$; (1) relatively high $T_{\rm CO}$ and (2) the presence of 
structural transition at $T_{\rm S}$ above $T_{\rm CO}$ in $R$BaMn$_{2}$O$_{6}$ with the small 
ionic size of $R^{3 + }$.}
\end{figure}

It is very interesting to compare Fig.~\ref{fig:fig4} with Fig. 1. Figure 4 is similar to 
Fig. 1 as a whole. There exist the characteristic phases such as the FM 
phase and the CE-type CO phase in both phase diagrams. In Fig.1, the FM 
phase appears in La$_{0.5}$Sr$_{0.5}$MnO$_{3}$ with $f$ = 0.996 (the average 
ionic size of La$^{3 + }$(1.36 {\AA})/Sr$^{2 + }$(1.44 {\AA}) = 1.40 {\AA}). 
In Fig.4, on the other hand, the FM phase appears around $R$ = La and the CO 
state becomes dominant for Nd and later rare earth. Incidentally if a 
hypothetical tolerance factor $f$' were calculated from the average ionic size 
of $R^{3 + }/$Ba$^{2 + }$, the FM phase would appear around $f$' = 1.026 
(LaBaMn$_{2}$O$_{6})$ far beyond $f$' = 1 and the CO phase would be stable 
around $f$' = 1 (from $f$' = 1.005 for SmBaMn$_{2}$O$_{6}$ to $f$' = 0.995 for 
YBaMn$_{2}$O$_{6})$. In $R$BaMn$_{2}$O$_{6}$, the MnO$_{2}$ sub-lattice is 
sandwiched by $R$O and BaO layers with the different lattice-sizes and as a 
result the MnO$_{6}$ octahedron itself is distorted in a peculiar manner 
that the oxygen atoms of MnO$_{2}$ square-lattice are strongly bound by 
$R^{3 + }$ resulting in a buckling of Mn and oxygen atoms in the MnO$_{2}$ 
square plane [7], in contrast to the rigid MnO$_{6}$ octahedra in ($R^{3 + 
}_{0.5}A^{2 + }_{0.5})$MnO$_{3}$. The ionic size of Ba$^{2 + }$ (1.61 
{\AA}) [6] is much larger than those of Sr$^{2 + }$ and all of $R^{3 + }$. In 
the combination of $R^{3 + }$/Ba$^{2 + }$ the mismatch between $R$O- and 
BaO-lattices is the smallest in La$^{3 + }$/Ba$^{2 + }$. Therefore lattice 
distortion is expected to be a little in LaBaMn$_{2}$O$_{6}$. Actually the 
structure of LaBaMn$_{2}$O$_{6}$ is tetragonal and the FM phase appears as 
the ground state. Here it should be noticed again that PrBaMn$_{2}$O$_{6}$ 
with the second smallest mismatch shows the FM to $A$-type antiferromagnetic 
metal (AFM) transition. A similar $A$-type AFM transition was previously 
reported in metal-disordered Pr$_{0.5}$Sr$_{0.5}$MnO$_{3}$ or 
Nd$_{0.5}$Sr$_{0.5}$MnO$_{3}$ with $f \sim $ 0.985 [8]. Among the FM phases 
the $T_{\rm C}$ is little dependent of the ratio, $r_{R^{3 + }}$/$r_{Ba^{2 + }}$. This 
suggests that the lattice distortion in $R$BaMn$_{2}$O$_{6}$ is not so 
considerable as the ferromagnetic interaction or DE interaction is 
influenced. Actually the Weiss temperature in the PM region above $T_{\rm S}$ is 
a ferromagnetic value about +300 K even in YBaMn$_{2}$O$_{6}$ which has the 
largest lattice distortion. On the other hand, the CO state becomes stable 
as the ratio, $r_{R^{3 + }}$/$r_{Ba^{2 + }}$ decreases. The $T_{\rm CO}$ increases across 
$T_{\rm C}$ around NdBaMn$_{2}$O$_{6}$ and reaches the champion record $T_{\rm CO}$ = 
480 K in YBaMn$_{2}$O$_{6}$. Our recent study of electron diffraction and 
neutron diffraction has revealed that YBaMn$_{2}$O$_{6}$ has a CE-type 
charge and orbital order with a 4-fold periodicity along the $c$-axis (2$\sqrt{2}$$a_{\rm p}$$\times$$\sqrt{2}$$b_{\rm p}$$\times$4$ c_{\rm p})$ [9]. Taking the layer-type metal-order 
into the consideration, this new type of charge and orbital order (4-CE-type 
CO) can be explained as follows: the orbital ordered pattern within the 
$a$-$b$ plane varies in the phase across the BaO- or YO-layer, that is $\alpha 
\alpha \beta \beta $-stacking along the $c$-axis, where the $\beta $-type 
is derived from the interconversion of the $d_{3x^{2} - r^{2}}$- and $d_{3y^{2} - 
r^{2}}$-sublattices in the ordinary CE-type layer ($\alpha $-type). 

There are remarkable features in the phase diagram of $R$BaMn$_{2}$O$_{6}$; (1) 
the high $T_{\rm CO}$ and (2) the presence of structural transition at $T_{\rm S}$ 
above $T_{\rm CO}$. It is easy to understand the relatively high $T_{\rm CO}$ in 
$R$BaMn$_{2}$O$_{6}$, because the absence of randomness at $A$-site and the 
layer-type metal-order are favorable for the charge ordering of Mn$^{3 + 
}$/Mn$^{4 + }$. Furthermore the phase diagram indicates that the increase of 
the mismatch between $R$O- and BaO-lattices also enhances the charge ordering. 
The structural transition at $T_{\rm S}$ is not accompanied by any charge and 
magnetic order but by the reduction of magnetic susceptibility. The 
temperature dependences of magnetic susceptibility suggest the change of 
magnetic interaction with Mn ions from ferromagnetic above $T_{\rm S}$ to antiferromagnetic 
below $T_{\rm S}$ [3]. Such novel transition is characteristic of the compounds 
with small ionic radii of $R^{3 + }$ in which the MnO$_{2}$ square-lattice is 
sandwiched by two rock-salt layers with extremely different lattice-sizes. 
This situation introduces a strong frustration to the MnO$_{2}$ sub-lattice 
and as a result the MnO$_{6}$ octahedron itself is heavily distorted leading 
to a complex structural deformation (triclinic or monoclinic) [7]. Such 
deformation must give a new perturbation to the competition of multi-degrees 
of freedom among charge, orbital, spin and lattice, and affect the 
characteristic properties such as CMR and charge/orbital ordering. We 
propose a possible orbital ordering, presumably $d_{x^{2} - y^{2}}$-type orbital 
ordering, at $T_{\rm S}$, referring to the $A$-type AFM. We have obtained some 
evidence for the orbital ordering from the detailed structural investigation 
by X-ray and neutron diffraction [7]. The freezing of the orbital, charge 
and spin degrees of freedom at the independent temperatures, $T_{\rm S}$, 
$T_{\rm CO}$ and $T_{\rm N}$, could be closely related to the peculiar structure of 
the metal-ordered perovskite-type manganites, that is a layer type and a low 
symmetric structure, an asymmetric distortion of MnO$_{6}$ octahedron and so 
on.

In the phase diagram of ($R^{3 + }_{0.5}A^{2 + }_{0.5})$MnO$_{3}$, the 
middle region ($f \sim $ 0.975) where the ferromagnetic DE and the 
antiferromagnetic CO interactions compete each other is responsible for 
various phenomena including CMR. Such region may correspond to the solid 
solution between NdBaMn$_{2}$O$_{6}$ and SmBaMn$_{2}$O$_{6}$ or 
PrBaMn$_{2}$O$_{6}$ in the phase diagram of $R$BaMn$_{2}$O$_{6}$. Actually 
rather complex behaviors have been observed in NdBaMn$_{2}$O$_{6}$ [5]. 
Furthermore our preliminary experiments have revealed the successful 
synthesis of metal-disordered manganites ($R^{3 + }_{0.5}$Ba$^{2 + 
}_{0.5})$MnO$_{3}$. The ($R^{3 + }_{0.5}$Ba$^{2 + }_{0.5})$MnO$_{3}$ 
will give us a chance of quantitative discussion on the effect of $A$-site 
randomness.

In summary, new metal-ordered perovskite-type manganites 
$R$BaMn$_{2}$O$_{6}$ ($R $= Y and rare earth elements) have been synthesized and 
investigated in the structures and electromagnetic properties. The 
$R$BaMn$_{2}$O$_{6}$ can be classified into three groups from the obtained 
structural and electromagnetic properties. The first group ($R$ = La, Pr and 
Nd) has a metallic ferromagnetic transition, followed by an $A$-type 
antiferromagnetic transition in PrBaMn$_{2}$O$_{6}$. The second group ($R$ = 
Sm, Eu and Gd) exhibits a CO transition, followed by an antiferromagnetic 
long range ordering. The third group ($R$ = Tb, Dy and Ho) shows successive 
three phase transitions, the structural, CO and magnetic transitions, as 
observed in YBaMn$_{2}$O$_{6}$. Comparing to metal-disordered ($R^{3 + 
}_{0.5}A^{2 + }_{0.5})$MnO$_{3}$, there are two remarkable features in 
$R$BaMn$_{2}$O$_{6}$; (1) the relatively high charge-order transition 
temperature and (2) the presence of structural transition without any charge 
and magnetic order above the charge order temperature in the third group. We 
propose a possible orbital ordering at the structural transition, that is a 
possible freezing of the orbital, charge and spin degrees of freedom at the 
independent temperatures in the third group. These are closely related to 
the structural feature that the MnO$_{2}$ sub-lattice is sandwiched by two 
kinds of rock-salt layers, $R$O and BaO with the different lattice-sizes and as 
a result the MO$_{6}$ octahedron itself is distorted in a peculiar manner.

The authors thank T. Yamauchi, M. Isobe, T. Matsushita, Z. Hiroi and H. 
Fukuyama for valuable discussion. This work is partly supported by 
Grants-in-Aid for Scientific Research (No. 407 and No. 758) and for Creative 
Scientific Research (No. 13NP0201) from the Ministry of Education, Culture, 
Sports, Science, and Technology.
\\
\\
\noindent\textbf{References}

\noindent[1] See for reviews: C. N. R. Rao and B. Raveau, \textit{Colossal Magnetoresistance, Charge Ordering and Related Properties of Manganese Oxides}; World Scientific, 
Singapore (1998).

\noindent[2] S. Mori, C. H. Chen, and S-W. Cheong, Phys. Rev. Lett. \textbf{81} 
(1998) 3972.

\noindent[3] T. Nakajima, H. Kageyama and Y. Ueda, J. Phys. Chem. Solids \textbf{63} 
(2002) 913.

\noindent[4] F. Millange, V. Caignaert, B. Domeng\`{e}s, and B. Raveau, Chem. 
Mater$.$ \textbf{10} (1998) 1974.

\noindent[5] in preparation

\noindent[6] R. D. Shannon, Acta Crystallogr. A\textbf{ 32} (1976) 751.

\noindent[7] T. Nakajima \textit{et al}., submitted to J. Solid State Chemistry.

\noindent[8] H. Kawano, R. Kajimoto, H. Yoshizawa, Y. Tomioka, H. Kuwahara and Y. 
Tokura, Phys. Rev. Lett. \textbf{78} (1997) 4253.

\noindent[9] H. Kageyama \textit{et al}., submitted to J. Phys. Soc. Japan.

\end{document}